\documentclass[conference, compsoc]{IEEEtran}
\usepackage{amsmath,amssymb,amsfonts}
\usepackage{spverbatim}
\usepackage{graphicx}
\usepackage{hyperref}
\usepackage{xurl}

\usepackage{refcount}
\newcounter{fncntr}
\newcommand{\fnmark}[1]{\refstepcounter{fncntr}\label{#1}\footnotemark[\getrefnumber{#1}]}
\newcommand{\fntext}[2]{\footnotetext[\getrefnumber{#1}]{#2}}

\renewcommand{\paragraph}[1]{\textit{#1}}

\begin {document}
\title{Per-sender neural network classifiers for email authorship validation}

\author{\IEEEauthorblockN{Rohit Dube}
\IEEEauthorblockA{\textit{Independent Researcher} \\
California, USA \\
}
}

\maketitle

\begin{abstract}
Business email compromise and lateral spear phishing attacks are among modern organizations' most costly and damaging threats.
While inbound phishing defenses have improved significantly, most organizations still trust internal emails by default, leaving themselves vulnerable to attacks from compromised employee accounts.
In this work, we define and explore the problem of authorship validation: verifying whether a claimed sender actually authored a given email.
Authorship validation is a lightweight, real-time defense that complements traditional detection methods by modeling per-sender writing style.
Further, the paper presents a collection of new datasets based on the Enron corpus. These simulate inauthentic messages using both human-written and large language model-generated emails.
The paper also evaluates two classifiers---a Naive Bayes model and a character-level convolutional neural network (Char-CNN)---for the authorship validation task.
Our experiments show that the Char-CNN model achieves high accuracy and F1 scores under various circumstances.
Finally, we discuss deployment considerations and show that per-sender authorship classifiers are practical for integrating into existing commercial email security systems with low overhead.
\end{abstract}

\begin{IEEEkeywords}
lateral spear phishing, business email compromise, email security, natural language processing, convolutional neural networks, stylometry
\end{IEEEkeywords}

\section{Introduction} \label{sec: intro}
Email-borne attacks remain a major cybersecurity attack vector.
``Phishing'' is a category of email attacks where the recipient of an email is tricked into divulging sensitive data or downloading malware
  \cite{ciscophishing}.
``Spear phishing'' is a sub-category of phishing where the email-based attack is targeted towards an individual or a small group of related individuals
  (usually by association with the same organization) \cite{ciscospear}.
``Business email compromise (BEC)'' is a sub-category of spear phishing where the attacker attempts to steal money from an organization \cite{ibmbec}.

Another sub-category of spear phishing is ``lateral spear phishing'' where an adversary uses a compromised account to send phishing emails to 
  a new set of recipients in the same organization \cite{ho2019detecting}.

Finally, one can imagine lateral BEC, where an attacker with access to a compromised email account attempts financial fraud by
  instructing the organization's employees to conduct legitimate-sounding actions that are, in fact, malicious. \fnmark{lateral-bec}
The employees may be fooled by the implicit authority of an internal email account and unknowingly enable the fraud.

\fntext{lateral-bec}{
  Research literature does not reference lateral BEC.
  However, given the prevalence of lateral spear phishing and BEC, lateral BEC almost certainly exists and is a subset of lateral spear phishing.
  The lack of public documentation is likely due to privacy and legal concerns in the aftermath of BEC attacks.
}

These lateral attacks are challenging to detect because many organizations do not pass internal emails through an email security system \cite{intermedia}.
Sometimes, this is done to minimize the cost of operating the email security system—costs increase as the volume of email through the system increases.
Other times, organizations do not perceive enough benefit to justify the operational complexity of scrutinizing internal emails \cite{conversation}.
In yet other cases, organizations' email infrastructure does not support the insertion of an email security stack for internal emails \cite{exchange}. \fnmark{colab}

\fntext{colab}{
 The problem stretches beyond email to other organizational collaboration tools \cite{mimecast2023}.
}

Central to the lateral attacks is the attempt by an attacker to impersonate an organization's employee.
Such attacks could be detected if a system were in place that checked internal emails for signs of impersonation, account takeover, and malicious content \cite{wendt2024ai}.

In this paper, we primarily focus on detecting impersonation using the content of email messages.
Towards our goal of detecting impersonation, we define an \textit{authorship validation task: this is the task of verifying whether a document (email)
  was authored by a specific author (the email account owner).}

While authorship attribution has been studied extensively in the context of identifying an author from a set of candidates, our focus is narrower and more practical:
  \textit{authorship validation} asks whether a given sender actually wrote a specific email.
This binary classification task aligns more directly with the needs of commercial email security systems, where real-time decisions must be made about individual messages.
Unlike general phishing detection, which typically looks for signs of malicious content or structure, authorship validation aims to detect subtle impersonation—even when the email contains no malware, links, or urgency cues.
To our knowledge, this task has not been formally defined or evaluated for real-world internal email communication.
Our work fills this gap by treating authorship as a behavioral signal that can be modeled and verified using sender-specific language patterns.

Our focus on impersonation notwithstanding, we advocate for a comprehensive sender-side detection stack that takes into account multiple signals,
  not just those in the email content.

The main contributions of this paper are:
\begin{itemize}
    \item \textbf{Definition of the authorship validation task}, a focused and practical variant of authorship attribution tailored for email security systems.
    This task enables real-time detection of impersonation using only the content of an email and a sender-specific behavioral model.
    \item \textbf{Construction of multiple datasets for authorship validation}, using the Enron email corpus as a base.
    \item \textbf{Comparative evaluation of two classifiers}---a Naive Bayes model and a character-level convolutional neural network (Char-CNN)---on these datasets.
    \item \textbf{Proposal of a scalable deployment architecture}, where per-sender neural network classifiers and behavioral profiles are integrated into commercial email security systems as part of a modular and extensible detection stack.
\end{itemize}

The rest of the paper is organized as follows.
Section~\ref{sec: related} discusses prior research.
Section~\ref{sec: threat-model} lays down the threat model assumed in this work.
Section~\ref{sec: dataset} describes the construction of datasets for authorship validation.
Section~\ref{sec: classifier} details the Naive Bayes and Char-CNN classifiers used for authorship validation.
Section~\ref{sec: results} and \ref{sec: analysis} report and analyze the performance of the two classifiers on the constructed datasets.
Section~\ref{sec: system} describes how authorship validation classifiers may be incorporated into commercial email security systems.
Section~\ref{sec: conclusion} summarizes our findings.
Finally, Section~\ref{sec: future} describes some ideas for continued exploration of authorship validation that are beyond the scope of this paper.

\section{Related work} \label{sec: related}
We review prior research work in three related fields: the use of sender profiles in detecting email-borne attacks, stylometry and authorship attribution, and publicly available descriptions of commercial email security systems.
Our review reveals space for using per-sender neural network-based classification techniques in modern-day email security systems.

\subsection {Sender profiles}
A few papers describe the value of creating and maintaining sender profiles to detect spear phishing and BEC attacks.

\cite {stringhini2015ain} explores sender-side detection of BEC emails (referred to as ``spearphishing'' in the cited paper).
The paper's authors describe a manual mechanism to build and maintain a behavioral profile per email sender in an organization.
A profile captures a sender's email habits with respect to a pre-determined list of attributes, including characters, words, and universal resource locators (URLs) in the body of the email,
  email formatting preferences (bullets, lists, paragraphs, signatures), email send times, and email recipients.
When it receives a new email from a sender, the email security system extracts the above-mentioned features.
Subsequently, the system feeds the features and the stored user profile to an ML, but a non-neural network classifier.
In turn, the classifier provides a verdict regarding the new email's authenticity compared to the sender's stored profile.
Our paper can be thought of as a modern neural-network-based extension of the approach in \cite {stringhini2015ain}.

\cite{duman2016emailprofiler} describes a mechanism to create sender profiles from email headers and bodies.
Relevant features are manually extracted from the emails.
Subsequently, an ML algorithm is employed to determine whether a new email is indeed from the actual sender or an attacker masquerading as the sender.
Note that the profiles are sender profiles but are created and deployed on the receiving end.
Using the profiles does not prevent emails from leaving the sending organization's infrastructure.

\cite{maleki2019behavioral} proposes stopping BEC emails by detecting them before they are sent.
The detection of BEC emails depends on manually pre-identified features extracted from each email.
However, \cite{maleki2019behavioral} proposes calculating centroids from extracted features.
A centroid represents the average email behavior of a user with respect to a feature.
Together, these centroids represent a user's email-sending behavior profile.
When a user's email deviates too far from the behavior represented by the centroids, the email is flagged for further analysis or blocking.

\cite{vorobeva2021detection} also discusses extracting various pre-determined features from emails to run ML classifiers against.
The idea remains to ascertain whether a sender actually authored an email.
The system described in \cite{vorobeva2021detection} implicitly maintains a sender profile but does not indicate whether the profile is kept at the sender or the receiver end.

The papers reviewed in this section have a few things in common:
\begin{itemize}
    \item Sender behavior profiles are determined from (manually) predefined features of a collection of emails authored by a sender.
    \item Email authorship is determined using an algorithm that takes the predefined features as input. Neural networks are not used.
\end{itemize}

We observe that with the availability of mature neural network frameworks and libraries, it may be possible to reduce or eliminate the need for manual feature extraction from emails \cite{pytorch}.

\subsection{Stylometry and authorship attribution}

Stylometry studies measurable patterns of writing style to characterize authors or distinguish between them.
Early work demonstrated that stable statistical properties of language can be used to resolve disputed authorship~\cite{mosteller1964inference}.
This analysis showed that function-word frequencies and other low-level stylistic cues persist across topics.
These findings established a foundational statistical framework for later research in computational authorship analysis.

Distance-based stylometric methods were later formalized through the introduction of the Delta measure~\cite{burrows2002delta}.
Delta quantifies stylistic distance using normalized word-frequency profiles and became widely adopted due to its simplicity.
Subsequent work showed that character n-grams often outperform word-level features in capturing stylistic regularities.
This result is particularly relevant for email, which is typically short, informal, and noisy.

A comprehensive survey of modern authorship attribution methods is provided in~\cite{stamatatos2009survey}.
The survey organizes prior work by feature types, learning paradigms, and evaluation settings.
It emphasizes that stylometric features are largely content-independent and complementary to semantic signals.
These properties make stylometry well suited for security and forensic applications.
Practical and computational aspects of authorship attribution are further examined in~\cite{koppel2009computational}, including robustness across domains and adversarial settings.

Within this broader stylometric context, several papers focus specifically on the authorship attribution problem.
Authorship attribution aims to determine the author of a given piece of writing and can be formulated as either an open-class or closed-class problem.
Neural approaches have become increasingly prominent in this area, particularly those based on character-level modeling.
Convolutional neural networks have been shown to effectively capture stylistic patterns from character-level signals~\cite{ruder2016character}.
Such models have also been extended to combine character and word representations to jointly model stylistic and topical cues.

Authorship attribution of short texts has received particular attention.
Character n-gram CNNs have been successfully applied to short and noisy domains such as social media~\cite{shrestha2017convolutional}.
Subsequent work further explored neural architectures for character- and word-level modeling to address the lack of explicit stylistic ``signature'' features in short texts~\cite{maneriker2021sysml}.
More recent surveys review a broad range of authorship-related problems, including attribution and large language model (LLM) generated text detection~\cite{huang2025authorship}.

While closely related, authorship attribution addresses a more difficult problem than the one considered in this paper.
Authorship attribution seeks to identify one author among many, whereas authorship validation focuses on verifying whether a single claimed author wrote a given document.
Authorship validation can therefore be framed as a binary classification problem.
This simplification reduces computational cost and makes it feasible to train sender-specific models for commercial email security systems.

Our work falls within the stylometric tradition by applying writing-style analysis to email security.
We model per-sender writing style rather than message content, leveraging character-level representations that are widely recognized as effective stylometric features.
By framing authorship validation as a behavioral signal for detecting impersonation and fraudulent messages, this work extends stylometric techniques to a practical email security setting.

\subsection {Commercial email security systems}
Two publicly available papers provide insight into the design of email security systems.

\cite{cidon2019high} describes a commercial system that has two stages.
The first stage uses information other than the subject and body (such as the email header) to determine if an email is suspicious.
Suspicious emails are taken through a second stage that analyzes the subject and body of the email.
If one of the analyses in the second stage also finds an email to be suspicious, it is blocked.
The second stage includes a natural language processing technique for the subject and body of the email.
The system creates a vector for each processed email using ``term frequency-inverse document frequency'' (TF-IDF).
A pre-trained classifier then classifies the vectors as either malicious or benign.

\cite{brabec2023modular} describes its authors' experience building, bootstrapping, and maintaining a commercial email security system that detects BEC attacks.
The system described breaks each incoming email into sentences or sentence fragments.
Each of these fragments is fed into statistical and AI models called detectors.
There are $\approx 90$ such detectors in the system, and each detector detects the presence of one or more specific behaviors in the fragment (e.g., urgency, call to action, frequent communication).
The output of the individual detectors is aggregated using logistic regression to create a final verdict.

A reading of these two papers indicates that:
\begin{itemize}
    \item Commercial systems exhibit a modular design so that additional ``signals'' can be integrated at a manageable design and implementation cost.
    \item The systems described do not explicitly maintain a sender profile and do not currently focus on stopping malicious emails from being sent (i.e., detection is on the receiving end).
\end{itemize}

\section{Threat Model} \label{sec: threat-model}

This section defines the threat model assumed throughout this paper and makes explicit the security assumptions that motivate the authorship validation task.
The structure of the threat model follows established threat modeling practices that emphasize assets, adversary capabilities, system boundaries, and defensive objectives \cite{shostack2014threat,shevchenko2018threat}.

We consider an enterprise email environment in which users primarily trust internal email messages by default.
We assume that standard email authentication mechanisms such as SPF, DKIM, and DMARC are already deployed and correctly enforce sender domain authenticity.
As a result, the attacker cannot spoof external sender identities but can send emails from a legitimate internal account, a setting commonly assumed in prior email phishing threat models \cite{olivo2013obtaining}.

The primary adversary is an attacker who has obtained access to a valid internal email account through credential compromise, malware, or account takeover.
The attacker’s goal is to impersonate the legitimate account owner in order to conduct lateral spear phishing or lateral business email compromise attacks.
We assume the attacker seeks to exploit implicit trust relationships within the organization rather than relying on malware attachments or malicious links.

The attacker may send short, benign-looking emails that request sensitive actions, financial transfers, or confidential information.
We assume the attacker has access to a limited subset of prior emails sent from the compromised account.
This limitation may arise because large-scale mailbox exfiltration or prolonged access could trigger email security systems, data loss prevention mechanisms, or anomaly detection controls.

We further assume the attacker may use LLMs or other automated tools to generate emails that are grammatically correct and contextually appropriate.
However, we assume the attacker can only condition such models on a relatively small number of authentic emails to avoid transferring large volumes of data or files.
As a result, the attacker’s ability to closely mimic the legitimate sender’s writing style is imperfect and constrained.

The attacker’s capabilities include crafting messages that avoid common phishing indicators such as urgency cues, suspicious URLs, or obvious social engineering language.
We assume the attacker does not have direct access to the parameters or internal state of the authorship validation classifier.
We assume the attacker may adapt over time by observing which emails are blocked or flagged, but does not receive explicit classifier feedback beyond delivery success or failure.

From a machine learning security perspective, we assume a black-box evasion threat model in which the adversary can probe the system indirectly through email delivery outcomes but cannot inspect or manipulate model internals \cite{papernot2018sok}.
We assume that attacks targeting the confidentiality of model parameters or the integrity of the training pipeline are out of scope.

The primary asset protected by the system is the integrity of sender identity in internal email communication.
A secondary asset is the trust relationship between employees that enables routine business operations, which has been identified as central to email-based attacks in prior threat modeling studies \cite{olivo2013obtaining}.

The system boundary includes the email subject and body content, which are the only inputs used by the authorship validation classifier.
Email metadata, headers, attachments, and network-level signals are considered out of scope for the classifier studied in this paper.

We assume that the classifier operates as one signal within a larger, modular email security detection stack.
The classifier’s output is not assumed to be a final enforcement decision, but rather an input into a broader aggregation framework, consistent with layered defense principles \cite{shostack2014threat}.

We explicitly scope our threat model to impersonation detection rather than intent classification.
As a result, emails that are authentically authored but malicious in intent are considered out of scope.
Similarly, emails that are inauthentic but benign in intent are still considered failures of sender identity integrity.

We assume the classifier is trained on historical emails authored by the legitimate sender and periodically retrained as new emails are sent.
We assume the training data is predominantly clean and correctly labeled, though occasional mislabeled samples may occur.

We do not consider targeted data poisoning attacks against the training pipeline in this work.
We do not consider denial-of-service attacks against the email infrastructure or classifier availability.
The threat model assumes realistic operational constraints, including limited per-sender training data and modest computational budgets, in line with practical threat modeling guidance \cite{shevchenko2018threat}.

Under this model, the defender’s objective is to detect deviations from a sender’s established writing style that indicate possible impersonation.
By making these assumptions explicit, the threat model clarifies the scope, limitations, and applicability of the proposed authorship validation system.

\section{Dataset construction} \label{sec: dataset}
We base our experiments on the Enron email corpus comprising email communications from Enron corporation's employees \cite{enron2015}, \cite{minkov2008activity}.
The Enron corpus is one of the few large-scale, real-world corpora of email communication made publicly available. It was initially released during legal investigations into Enron Corporation.
The corpus contains email accounts from around 150 employees, but most messages come from a much smaller subset of key users—mainly senior executives and managers.

\subsection{Authentic emails}
Two authentic email sets are manually derived from the Enron corpus (see Table \ref{tab: authentic}).

\begin{table}[htbp]
    \caption {\label{tab: authentic} Authentic email sets}
    \begin{center}
            \begin{tabular}{|c|c|c|}
            \hline
            \textbf{Set} & \textbf{Description} & \textbf{\# emails} \\ \hline
            Authentic-1 & Sender \texttt{kaminski-v} & 369 \\ \hline
            Authentic-2 & Sender \texttt{stclair-c} & 369 \\ \hline
            \end{tabular}
    \end{center}
\end{table}

\subsubsection {Sender \texttt{kaminski-v}} \label{sec: kaminski-v}
We selected emails from the ``sent'' folder of the randomly chosen sender \texttt{kaminski-v}.
This sender is one of the top-12 senders in the corpus and has authored over 3000 emails, providing a rich source of human-authored content.
To simulate the behavior of a moderately active email user, we limited our selection to the first 600 email files in this folder (numbered 1 through 600, inclusive).

Each email in this set was manually cleaned to remove non-authored content.
Specifically, we removed the email headers, eliminated quoted content from earlier messages in the thread, and discarded any email containing only a forwarded message without additional commentary.
The resulting emails represent original authored content and are treated as authentic.
These emails are implicitly assigned the label \( y = 1 \), corresponding to human-authored, authentic messages.
Table \ref{tab: kaminski-v} shows some statistics from our chosen sender.

Most emails in this cleaned set represent day-to-day business activity.
Many are short, and none exceed 5120 characters (including the length of the subject line).

\begin{table}[htbp]
    \caption {\label{tab: kaminski-v} Statistics for sender \texttt{kaminski-v}}
    \begin{center}
            \begin{tabular}{|c|c|}
            \hline
            \textbf{Description} & \textbf{\# emails} \\ \hline
            Total email files in kaminski-v/sent folder & 3462 \\ \hline
            Email files between 1. and 600. & 596 \\ \hline
            Cleaned (authentic) email files & 369 \\ \hline
            \end{tabular}
    \end{center}
\end{table}

\subsubsection {Sender \texttt{stclair-c}}
Using the same procedure as that for \texttt{kaminski-v}, we obtain 369 emails from a new randomly chosen Enron sender: \texttt{stclair-c}.
This sender is not one of the top 12 senders in the Enron corpus.

While we would like to have a diverse set of senders, the manual labor involved in extracting the subject and body of individual emails constrains our effort.
Even so, having two distinct randomly chosen senders provides us with some assurance of the generalizability of subsequent results.

\subsection{Inauthentic emails}
Obviously, authentic emails by themselves are insufficient to experiment with authorship validation.
We need emails similar to the cleaned ones in the authentic sets, but written in a different style or by different authors.
With two classes of emails, we hope to construct classifiers that can distinguish between them.

We created three inauthentic sets (see table \ref{tab: inauthentic}).
The first one is generated using an LLM with the authentic set as input.
The second is a mixture of emails from other Enron senders.
The third is a mixture of the previous two inauthentic sets with third-party-sourced BEC emails added in.
The three sets represent increasing complexity in the inauthentic emails.

\begin{table}[htbp]
    \caption {\label{tab: inauthentic} Inauthentic email sets}
    \begin{center}
            \begin{tabular}{|c|c|c|}
            \hline
            \textbf{Set} & \textbf{Description} & \textbf{\# emails} \\ \hline
            Inauthentic-1 & LLM-generated & 369 \\ \hline
            Inauthentic-2 & Other top senders & 369 \\ \hline
            Inauthentic-3 & Inauthentic-1 & 1017 \\
                          & + Inauthentic-2 & \\
                          & + third-party BEC & \\ \hline
            \end{tabular}
    \end{center}
\end{table}

\subsubsection{LLM-generated inauthentic set} \label {sec: inauthentic-llm}
For the first inauthentic set (Inauthentic-1), we used an LLM (GPT-4) to generate a stylistic variant of each cleaned email in Authentic-1.
The idea is inspired by \cite{dube2025building}, where an LLM is used to create a synthetic BEC dataset.

\begin{figure} [htbp]
    \begin{center}
    \includegraphics[scale=0.30]{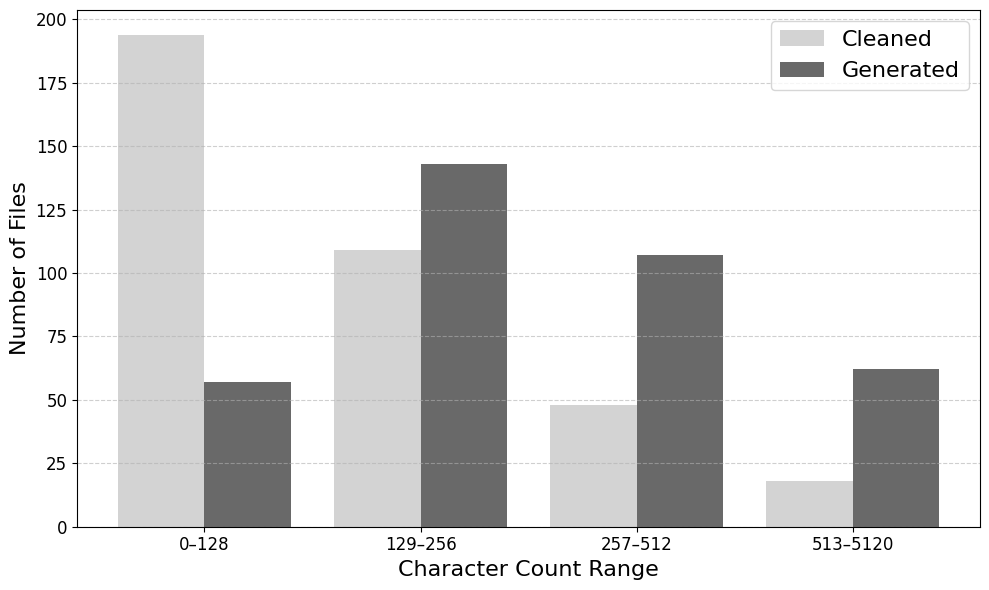}
    \caption {\label{fig: histogram} Email length distribution: authentic (clean original) versus inauthentic (generated) emails}
    \end{center}
\end{figure}

The LLM was prompted to rewrite the content in the style of the \texttt{New York Times}, producing more polished output while preserving the original's core content and intent \cite{openai}. \fnmark{openai}
On average, the generated emails are longer and more formally written than the original cleaned emails (see Figure \ref{fig: histogram}).
These generated emails are treated as inauthentic and are implicitly assigned the label \( y = 0 \).

\fntext{openai} {
  The model \texttt{gpt-4-turbo} and the prompt \textit{Rewrite the following text in the tone of the New York Times} were used via an application programming interface.
  Switching from \texttt{New York Times} to other well-known news outlets such as the \texttt{British Broadcasting Corporation} did not yield substantially different output.
}

Appendix \ref{sec: clean} shows an example of a clean email from the dataset.
Appendix \ref{sec: generated} shows the corresponding generated email.
Most readers will notice the two emails' differences in style and tone, even though the main point is identical.

We experimented with generating inauthentic emails by supplying the LLM with a few authentic email samples and asking it to create hundreds of variants.
While this idea is viable, the chosen method of inauthentic email construction was better at guaranteeing an adequate number of emails written in a different style than the original.

\subsubsection {Other top senders' emails as the inauthentic set} \label {sec: inauthentic-enron}
For the second inauthentic set (Inauthentic-2), we take 369 email samples from 11 of the 12 top senders in the Enron corpus (excluding \texttt{kaminski-v} whose emails are in Authentic-1).
Thus, the emails in the inauthentic set are human-authored, just like those in the authentic sets.

Note that the emails in Inauthentic-2 are cleaned per the same tedious process described above in Section \ref{sec: kaminski-v}.

\subsubsection {Mixture of emails as the inauthentic set} \label {sec: inauthentic-mix}
For the third inauthentic set (Inauthentic-3), we mixed the 369 LLM-generated emails in Section \ref{sec: inauthentic-llm} (Inauthentic-1) with the 369 emails from the other top senders in Section \ref{sec: inauthentic-enron} (Inauthentic-2), and an additional 279 BEC emails from a third-party \cite{dube2025building}, for a total of 1,017 inauthentic emails.
Here, the idea is to simulate a heterogeneous combination of emails.

\subsection {Creating datasets from authentic and inauthentic email sets} \label {sec: four-datasets}
Finally, we take different combinations of authentic and inauthentic email sets to create datasets on which authorship validation classifiers can be tested.

The first dataset (Dataset-1) consists of an equal number of authentic and inauthentic emails.
The authentic emails are from Authentic-1, and the inauthentic emails are from the corresponding LLM-generated Inauthentic-1.
Note that in Dataset-1, a generated email is not guaranteed to be in the same set (train or test) as the clean original email from which it was generated.
This is because each email is treated independently during the random split between train and test.

We create a second dataset (Dataset-2) that consists of the authentic emails from Authentic-1 and the inauthentic emails from Inauthentic-2.
This dataset is more challenging for an authorship validation classifier than Dataset-1, as the inauthentic emails are from multiple senders and thus written in multiple different styles.

Similarly, we create a third dataset (Dataset-3) that has the same authentic emails as above and inauthentic emails from Inauthentic-3.
Dataset-3 is even more challenging for a classifier than Dataset-2, as the inauthentic set has emails from multiple senders and multiple sources.

Finally, we create a fourth dataset (Dataset-4) by varying the authentic set (switching from Authentic-1 to Authentic-2) while keeping the inauthentic set (Inauthentic-3) in place.

Table \ref{tab: datasets} lists the component authentic and inauthentic email sets for each of the four datasets constructed.
All four datasets are used for both training and evaluation of text classifiers.
While the $80 : 20$ split between training and test sets is determined dynamically at runtime, it is implemented deterministically to ensure the replicability of experimental results.

\begin{table}[htbp]
    \caption {\label{tab: datasets} Datasets created for classifier experiments}
    \begin{center}
            \begin{tabular}{|c|c|c|c|}
            \hline
            \textbf{Dataset} & \textbf{Authentic} & \textbf{Inauthentic}& \textbf{\# emails} \\ \hline
            Dataset-1 & Authentic-1 & Inauthentic-1 & 738 \\ \hline
            Dataset-2 & Authentic-1 & Inauthentic-2 & 738 \\ \hline
            Dataset-3 & Authentic-1 & Inauthentic-3 & 1386 \\ \hline
            Dataset-4 & Authentic-2 & Inauthentic-3 & 1386 \\ \hline
            \end{tabular}
    \end{center}
\end{table}

\section {Classifiers for authorship validation} \label{sec: classifier}
We train two classifiers to assess the authenticity of email messages using the datasets described above in Section \ref{sec: dataset}.

The first is a Naive Bayes classifier, which serves as a simple and interpretable baseline \cite{mccallum1998comparison}.
We do not expect this classifier to have outstanding performance, but we do expect to gain insights into writing styles from the classifier.

The second is a character-level convolutional neural network employed to evaluate whether increased model complexity yields better performance on our modestly sized datasets \cite{lecun1998gradient}, \cite{zhang2015character}.
Using a character-level convolutional neural network is inspired by \cite{shrestha2017convolutional} that successfully uses character n-grams for authorship attribution of short texts.
We surmise that a classifier that has worked well for authorship attribution should work as well or better for authorship validation.

We describe each chosen classifier in detail below.
Code corresponding to the classifiers is available at \cite{dube2025snn}.

\subsection {Naive Bayes} \label{sec: nb}
The Naive Bayes algorithm is a probabilistic classifier based on Bayes' Theorem, assuming conditional independence between features given the class label.
For a document represented as a vector of features \( \mathbf{x} = (x_1, x_2, \ldots, x_n) \), the algorithm estimates the posterior probability of a class \( c \in \{c_1, c_2\} \) as
\begin{equation}
    P(c \mid \mathbf{x}) \propto P(c) \prod_{i=1}^n P(x_i \mid c),
\end{equation}
where \( P(c) \) is the prior probability of class \( c \), and \( P(x_i \mid c) \) is the likelihood of feature \( x_i \) given the class.

In our implementation, features correspond to word counts.
The model parameters are estimated using maximum likelihood.
Given that our task is to verify authorship for a specific individual, we restrict our classification problem to two classes (\( c_1 = \text{authentic} \), \( c_2 = \text{inauthentic} \)).

Despite its simplifying assumptions, Naive Bayes performs well on text classification tasks due to its robustness and computational efficiency.
It thus provides a valuable benchmark for more complex models.

\subsection{Character-Level Convolutional Neural Network (Char-CNN)} \label{sec: char-cnn}

We employ a character-level convolutional neural network (Char-CNN) to determine whether a given document was authored by a specific individual.
The task is formulated as binary classification, where a label of \(1\) indicates authorship by the target individual and a label of \(0\) indicates non-authorship.

\paragraph{Input Representation}
Each document is represented as a fixed-length sequence of character indices drawn from a predefined vocabulary.
Documents longer than \(L = 1024\) characters are truncated, and shorter documents are padded with a special padding symbol.
The resulting input tensor is defined as:
\begin{equation}
    \mathbf{X} \in \mathbb{N}^{B \times L},
\end{equation}
where \(B = 32\) denotes the batch size.

\paragraph{Embedding Layer}
Character indices are mapped to dense vector representations using a learned embedding matrix:
\begin{equation}
    \mathbf{E} = \text{Embedding}(\mathbf{X}) \in \mathbb{R}^{B \times L \times d},
\end{equation}
where \(d = 16\) is the embedding dimension.
The embedding tensor is permuted to
\begin{equation}
    \mathbf{E}' \in \mathbb{R}^{B \times d \times L},
\end{equation}
to match the input format expected by one-dimensional convolutional layers.

\paragraph{Convolutional Layers}
We apply three parallel one-dimensional convolutional layers with kernel sizes \(k \in \{7, 5, 3\}\).
Each convolution produces \(F = 64\) feature maps.
For each kernel size \(k_i\), the convolutional transformation is defined as:
\begin{equation}
    \mathbf{C}_i = \text{ReLU}\!\left(
        \text{BatchNorm}_i\!\left(
            \text{Conv1D}_i(\mathbf{E}')
        \right)
    \right),
\end{equation}
where batch normalization is applied to the convolutional outputs prior to the non-linear activation.
This ordering stabilizes the distribution of intermediate activations and facilitates optimization.

\paragraph{Pooling}
Each convolutional output is passed through adaptive max pooling over the temporal dimension:
\begin{equation}
    \mathbf{P}_i = \text{MaxPool}(\mathbf{C}_i) \in \mathbb{R}^{B \times F}.
\end{equation}
Adaptive max pooling aggregates each feature map into a single scalar by selecting the maximum activation across the sequence dimension.

\paragraph{Feature Concatenation and Regularization}
The pooled feature vectors from all convolutional branches are concatenated and regularized using dropout:
\begin{equation}
    \mathbf{H} = \text{Dropout}\!\left(
        [\mathbf{P}_1; \mathbf{P}_2; \mathbf{P}_3]
    \right) \in \mathbb{R}^{B \times 3F},
\end{equation}
where a dropout rate of \(0.5\) is applied during training to reduce overfitting.

\paragraph{Output Layer}
The concatenated feature representation is passed through a fully connected layer to produce a scalar logit:
\begin{equation}
    \mathbf{z} = \mathbf{H}\mathbf{W} + \mathbf{b} \in \mathbb{R}^{B \times 1},
\end{equation}
where \(\mathbf{W} \in \mathbb{R}^{3F \times 1}\) and \(\mathbf{b} \in \mathbb{R}\).
No activation function is applied at this stage.

\paragraph{Training Objective}
The model is trained using the binary cross-entropy loss with logits:
\begin{equation}
    \mathcal{L}(\mathbf{y}, \mathbf{z}) =
    \frac{1}{B} \sum_{b=1}^{B}
    \Big[
        - y_b \log \sigma(z_b)
        - (1 - y_b) \log (1 - \sigma(z_b))
    \Big],
\end{equation}
where \(\sigma(\cdot)\) denotes the sigmoid function and \(\mathbf{y} \in \{0,1\}^{B \times 1}\) represents the ground-truth labels.
This formulation is implemented using \texttt{BCEWithLogitsLoss}, which improves numerical stability by combining the sigmoid operation with the loss computation.

\paragraph{Inference}
During evaluation, predicted probabilities are obtained by applying the sigmoid function to the model logits.
A document is classified as authored by the target individual if the predicted probability is greater than or equal to \(0.5\); otherwise, it is classified as non-authored.

\section {Experimental results} \label{sec: results}
Each of the four datasets described in Section \ref{sec: dataset} is split $80 : 20$ into training and test sets.
Subsequently, both the Naive Bayes and Char-CNN classifiers are trained on the training set, and the classifiers' performance is measured on the test set.
Tables \ref{tab: dataset1-results}, \ref{tab: dataset2-results}, \ref{tab: dataset3-results}, \ref{tab: dataset4-results} show the performance results.

The Naive Bayes classifier demonstrates solid performance, given its simplicity.
The Char-CNN classifier outperforms Naive Bayes across most reported metrics, likely by learning subtle stylistic patterns from raw character sequences, though at the cost of reduced interpretability.

\begin{table}[htbp]
    \caption {\label{tab: dataset1-results} Naive Bayes and Char-CNN performance over Dataset-1}
    \begin{center}
            \begin{tabular}{|c|c|c|}
            \hline
            \textbf{Metric} & \textbf{Naive Bayes} & \textbf{Char-CNN} \\ \hline
            Accuracy & 0.9459 & 0.9662 \\ \hline
            F1 (macro) & 0.9513 & 0.9654 \\ \hline
            True positives & 58 & 60 \\ \hline
            True negatives & 83 & 83 \\ \hline
            False positives & 3 & 3 \\ \hline
            False negatives & 4 & 2 \\ \hline
            \# training epochs & NA & 43 \\ \hline
            Training stopped early & NA & Yes \\ \hline
            \end{tabular}
    \end{center}
\end{table}

\begin{table}[htbp]
  \caption {\label{tab: dataset2-results} Naive Bayes and Char-CNN performance over Dataset-2}
  \begin{center}
          \begin{tabular}{|c|c|c|}
          \hline
          \textbf{Metric} & \textbf{Naive Bayes} & \textbf{Char-CNN} \\ \hline
          Accuracy & 0.9122 & 0.9892 \\ \hline
          F1 (macro) & 0.9115 & 0.9859 \\ \hline
          True positives & 61 & 60 \\ \hline
          True negatives & 74 & 86 \\ \hline
          False positives & 12 & 0 \\ \hline
          False negatives & 1 & 2 \\ \hline
          \end{tabular}
  \end{center}
\end{table}

\begin{table}[htbp]
  \caption {\label{tab: dataset3-results} Naive Bayes and Char-CNN performance over Dataset-3}
  \begin{center}
          \begin{tabular}{|c|c|c|}
          \hline
          \textbf{Metric} & \textbf{Naive Bayes} & \textbf{Char-CNN} \\ \hline
          Accuracy & 0.9101 & 0.9892 \\ \hline
          F1 (macro) & 0.8923 & 0.9859 \\ \hline
          True positives & 70 & 70 \\ \hline
          True negatives & 183 & 205 \\ \hline
          False positives & 22 & 0 \\ \hline
          False negatives & 3 & 3 \\ \hline
          \end{tabular}
  \end{center}
\end{table}

\begin{table}[htbp]
  \caption {\label{tab: dataset4-results} Naive Bayes and Char-CNN performance over Dataset-4}
  \begin{center}
          \begin{tabular}{|c|c|c|}
          \hline
          \textbf{Metric} & \textbf{Naive Bayes} & \textbf{Char-CNN} \\ \hline
          Accuracy & 0.8705 & 0.9856 \\ \hline
          F1 (macro) & 0.8530 & 0.9811 \\ \hline
          True positives & 73 & 69 \\ \hline
          True negatives & 169 & 205 \\ \hline
          False positives & 36 & 0 \\ \hline
          False negatives & 0 & 4 \\ \hline
          \end{tabular}
  \end{center}
\end{table}

\section {Analysis} \label{sec: analysis}
We consider the performance of Naive Bayes and Char-CNN over each of the four datasets in order.
The analysis supports the viability of Char-CNN for authorship validation in real-world systems.

\subsection {Performance over Dataset-1}
First, we analyze the performance of the two classifiers over Dataset-1 separately and in some detail.
Much of this analysis applies to the performance of the classifiers over the other three datasets as well.

\subsubsection{Naive Bayes}
\begin{figure} [htbp]
    \begin{center}
    \includegraphics[scale=0.60]{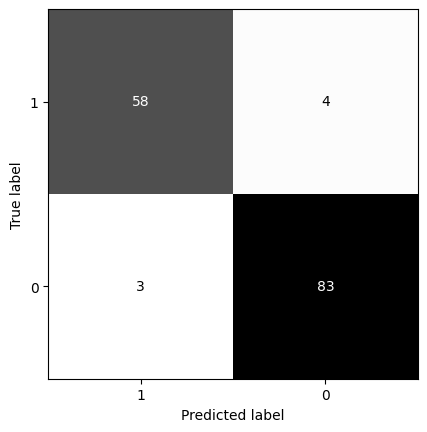}
    \caption {\label{fig: nb} Confusion matrix for Naive Bayes classifier}
    \end{center}
\end{figure}

The Naive Bayes classifier provides a reasonable baseline for the authorship validation task.
It achieves a macro-averaged F1 score of 0.9513, which reflects a solid performance given the model's simplicity and the dataset's limited size (see Table \ref{tab: dataset1-results}).
The classifier correctly identifies 58 authentic emails and misclassifies four as inauthentic (see Figure \ref{fig: nb}).
It also correctly classifies 83 inauthentic emails while mislabeling three as authentic.
The overall accuracy of 0.9459 indicates that the model captures some meaningful patterns.
As Naive Bayes does not involve iterative training, metrics such as training epochs and early stopping are not applicable.
While limited in expressiveness, Naive Bayes offers a valuable benchmark for assessing the added value of more complex models in the authorship validation task.

\begin{figure} [htbp]
    \begin{center}
    \includegraphics[scale=0.40]{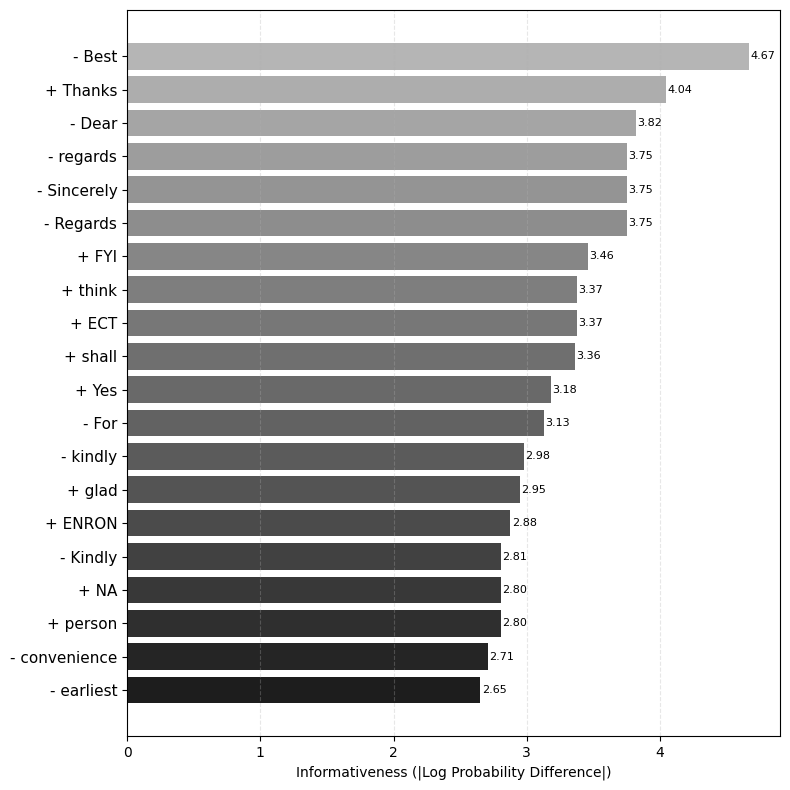}
    \caption {\label{fig: top-words} Top discriminative words in Naive Bayes classification}
    \end{center}
\end{figure}

To identify the most informative words for distinguishing between authentic and inauthentic emails, we analyzed the word-level log probability differences computed by the Naive Bayes classifier.
For each word \( w \), we computed the difference in log probabilities under the two classes: \( \log P(w \mid \text{authentic}) - \log P(w \mid \text{inauthentic}) \).
This quantity reflects how strongly a word favors one class over the other.
However, we took the absolute value of this difference to identify words that are simply informative regardless of class.
By ranking words according to \( \lvert \log \! \left( P\!\left(w \,\vert\, \text{authentic} \right) \right) - \log \! \left( P\!\left(w \,\vert\, \text{inauthentic} \right) \right) \rvert \),
we were able to surface terms that had the greatest discriminative power in the classification task.
Figure \ref{fig: top-words} illustrates our dataset's most informative words discovered through this method.

Words such as ``Thanks'', ``Dear'' ``Sincerely'' and ``FYI'' show high informativeness scores, suggesting that the original author used them very differently from the inauthentic variants.
In particular, the use of ``Thanks'' and ``FYI'' shows a degree of informality in the original author's writings.
On the other hand, using ``Dear'' and ``Sincerely'' underlines the more formal nature of the inauthentic emails.

\subsubsection{Char-CNN}
\begin{figure} [htbp]
    \begin{center}
    \includegraphics[scale=0.60]{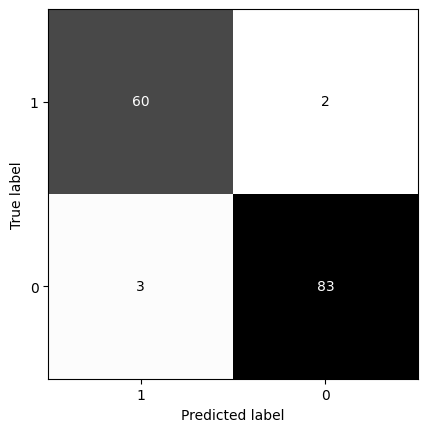}
    \caption {\label{fig: char-cnn} Confusion matrix for Char-CNN classifier}
    \end{center}
\end{figure}

\begin{figure} [htbp]
    \begin{center}
    \includegraphics[scale=0.45]{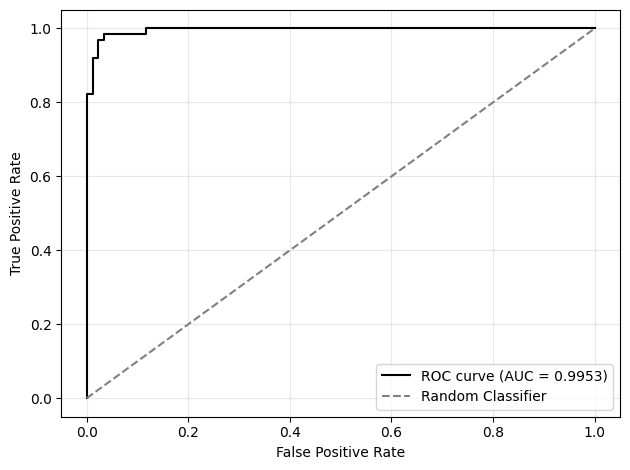}
    \caption {\label{fig: roc-auc} ROC curve for Char-CNN classifier performance}
    \end{center}
\end{figure}

The Char-CNN classifier also performed well on the authorship validation task, outperforming the Naive Bayes baseline across multiple metrics.
Despite being trained on a modest dataset, the Char-CNN classifier reached an accuracy of 0.9662 and a macro F1 score of 0.9654, compared to 0.9459 and 0.9513, respectively
  for Naive Bayes (see Table \ref{tab: dataset1-results} and Figure \ref{fig: char-cnn}).
The Char-CNN classifier correctly identified 60 authentic and 83 inauthentic emails, with only two false negatives and three false positives on the test set.
For completion, we show the corresponding receiver operating characteristic (ROC) curve in Figure \ref{fig: roc-auc}.

Appendix \ref{sec: fp} shows an example of one of the false positive email files.
While the email in the example is slightly more formal than the original, Char-CNN's interpretation of the formality does not overcome the brevity of the message.
Incidentally, Naive Bayes classified this email correctly.

Appendix \ref{sec: fn} shows an example of one of the false negative email files.
This email contains a moderate-sized subject line and a single URL in the body.
Given the modest dataset, it appears that Char-CNN did not have enough training examples to correctly adjudicate such emails.
Naive Bayes also failed to classify this email correctly.

One likely reason for the Char-CNN’s improved performance is its ability to learn hierarchical features from raw character sequences, which enables it to model subtle stylistic and structural cues in writing that may not be captured by word frequency statistics alone.
While Naive Bayes provides an interpretable framework based on word-level probabilities, its simplifying assumptions and limited expressiveness may constrain its effectiveness in distinguishing nuanced patterns.
In contrast, Char-CNN’s performance benefits from its greater representational capacity, but its internal behavior is significantly more complex to interpret, making it challenging to pinpoint which specific features the model relies on for classification.

\begin{figure} [htbp]
    \begin{center}
    \includegraphics[scale=0.35]{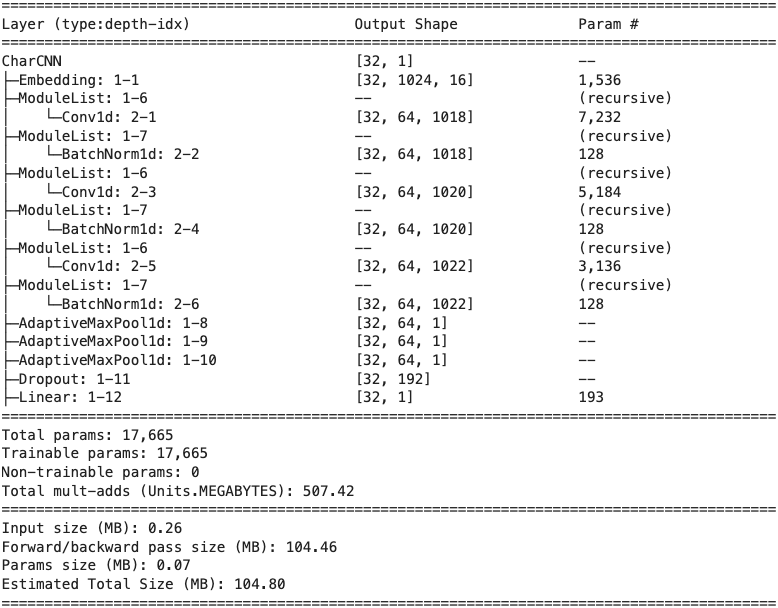}
    \caption {\label{fig: parameters} Trainable Parameters in Char-CNN}
    \end{center}
\end{figure}

Figure \ref{fig: parameters} enumerates the trainable parameters in Char-CNN.
The output is from \texttt{torchinfo}, a companion package to \texttt{pytorch} that was used to code the network \cite{pytorch}.
At $17,665$ parameters, the Char-CNN network is modestly sized at best.
Given the limited training data and model size, further improvements may be possible with a larger dataset and a more sophisticated architecture.

\subsection {Performance over Dataset-2}
As one can see from Table \ref{tab: dataset2-results}, the performance of the Naive Bayes classifier seems to have degraded moderately on Dataset2---there are now 12 false positives and one false negative compared to four and three, respectively, in the previous experiment.
In comparison, the performance of the Char-CNN classifier has improved modestly.
In fact, Char-CNN records zero false positives and two false negatives, compared to three and two, respectively, in the prior experimental run.

\subsection {Performance over Dataset-3}
The results in Table \ref{tab: dataset3-results} show Char-CNN (F1 score of 0.9859) clearly outperforming the Naive Bayes classifier (F1 score of 0.8923).
Naive Bayes performs somewhat more poorly on Dataset-3 compared to Dataset-2.
On the other hand, Char CNN's performance is stable across these two datasets.

\subsection {Performance over Dataset-4}
With Dataset-4, we seek to understand if the classifier performance observed thus far is somehow heavily dependent on the composition of the Authentic-1 email set.
Note that Dataset-4 switches in Authentic-2 for Authentic-1.

The results from an experimental run over Dataset-4 are in Table \ref{tab: dataset4-results}.
As in the previous experiments, Char-CNN outperforms the Naive Bayes classifier.
The performance of Naive Bayes slips further compared to the previous experiments.
However, the performance of Chair-CNN remains solid and is comparable to that in previous experimental runs.

\section {The case for authorship validation in commercial systems} \label{sec: system}
From Section \ref{sec: intro}, it is clear that a class of email-borne attacks—lateral spear phishing and lateral BEC—could be mitigated at the sender (potentially as the organization's email handling system receives the email).
From Section \ref{sec: related}, we know that there is an opportunity to use ML-based classifiers in commercial email security systems to detect whether a sender
  actually authored an email that comes from their email account.
From Sections \ref{sec: results} and \ref{sec: analysis}, we know that even simple classifiers can provide an independent authorship validation signal and
  that there is room for neural network-based classifiers to improve the quality of that signal.

How might the sender profile and authorship validation ideas presented above be incorporated into commercial email security systems?
In the rest of this section, we sketch an architecture that integrates authorship validation neural network classifiers into sender profiles.

\subsection {Sender profiles}
Commercial email security systems already maintain an implicit sender profile when they create relationship graphs to track email communication between senders and
  recipients \cite{cidon2019high}, \cite{brabec2023modular}.
We suggest that this sender profile be made explicit, expanded to the organization's internal email handling, and that other aspects of the sender's behavior be tracked.
The sender profile could capture information such as typical time-of-day, day-of-week characteristics, and the email client used for emails sent by a user \cite{stringhini2015ain}, \cite{duman2016emailprofiler}.
Making the sender profile extensible so that yet-to-be-discovered data and techniques can be inserted into it is merited.

\subsection {Per-sender neural network classifiers}
A neural network classifier, such as Char-CNN, could be one of the techniques tethered to a sender profile.

The general idea is to maintain a neural network classifier per sender in an organization.
The classifier architecture could be the same across all profiles (although this is not strictly necessary),
  but the parameters (weights) of the classifier are unique to each classifier instance.
The classifiers are trained on the emails sent and received by each sender.
Once trained, the classifiers provide an authorship validation signal to the email security system per sender.
Newly sent emails are batched and used to retrain the classifiers incrementally.

One might be concerned about the computing resources to train or retrain a classifier.
Training Char-CNN on Dataset-3 (1,108 emails in training set) takes 10 minutes and 55 seconds on an Apple M2 Pro with 16GB of RAM while other user processes run on the system.
Thus, this consumer-grade machine can support almost 4,000 training runs per month. \fnmark{calculation}
Enterprise server-grade machines would yield significantly higher monthly training runs, indicating that the additional computing resources needed for classifier training are modest.

\fntext{calculation}{
  Calculated as ($\frac{30 \times 24 \times 60 \times 60}{655}$), where 30 is the number of days in a month, 24 is the number of hours in a day, 60 is the number of minutes in an hour, and 60 the number of seconds in an hour.
  655 is the number of seconds for the sample training run mentioned.
}

The time needed to test (evaluate) 278 emails with Char-CNN on the same machine is 1.87 seconds.
As such, the runtime cost of using Char-CNN in a production environment is likely minor.

\subsection {Modular detection stack for sent emails}
The signals from the different components of the sender profile are aggregated to declare a new email from a sender authentic or inauthentic.
Given that such detection involves multiple signals, it makes sense to create a modular detection stack for emails being sent, similar to the one that we have for emails being received from other organizations \cite{stringhini2015ain}, \cite{duman2016emailprofiler}.
Further, the detection stack must be extensible to support new sender signals as they are discovered.

\subsection {Handling false positives and negatives}
A commercial email security system with authorship validation must also provide a mechanism for handling user-reported false positives and negatives.
Many organizations have already implemented user-facing tools that allow employees to flag emails marked incorrectly.
Authorship validation can be integrated into this system.
Once flagged, false positive and negative emails are reviewed---either manually or using automated analysis---to verify their true label.
For false positives, if the verification step confirms that the email is authentic, it is released to the recipient's inbox.
For false negatives, if the verification step confirms that the email is inauthentic, the email security team is alerted, necessitating further triage and containment.

Verified false positives and negatives can be incorporated into future model retraining, helping the system learn more accurate distinctions between authentic and inauthentic messages.
However, this must be done carefully---only emails flagged incorrectly by the validation classifier must be added back to the training set with the correct label.
Remember that the detection stack combines multiple signals---an email reported as a false positive or negative may, in fact, have been classified correctly by the validation classifier!

While cumbersome, integrating user feedback makes the system more robust and minimizes disruption to regular business communication.

\section {Conclusion} \label{sec: conclusion}
In this paper, we have focused on detecting impersonators carrying out lateral spear phishing and BEC attacks.
Toward such detection, we first defined the authorship validation task.
Then, we created research datasets of emails to conduct authorship validation.
For validation, we experimented with and analyzed the performance of two classifiers:
  a simple Naive Bayes classifier and a more sophisticated convolutional neural network.
Our experiments showed that authorship validation can provide a valuable signal
  to email security systems to detect impersonators.
Finally, we outlined how email security systems can incorporate classifiers built for authorship validation.

\section {Future work} \label{sec: future}
Several areas studied or referenced in the paper invite future work.
We aggregate these below for easy reference.

\textit{Dataset scale.}
While our experiments are based on the widely used Enron email corpus, our datasets are relatively modest as they are drawn from a few senders' communication histories.
The constraining factor is the amount of tedious manual labor required to create the datasets from the Enron corpus.
Future work should explore techniques such as crowd-sourcing to create larger, more diverse datasets.
In particular, creating a standard dataset for authorship validation, similar to equivalent datasets in other fields, would be useful to researchers \cite{dube2024faulty}.

\textit{Exploration of the classifier design space.}
This involves evaluating the performance of existing classifiers (including those used for authorship attribution) and the creation of new classifier designs optimized for authorship validation.
Classifiers such as very deep convolutional neural networks are of particular interest \cite{conneau2016very}.

\textit{Adversarial evasion and robustness.}
As with any ML-based system, adversarial actors may attempt to craft emails that evade detection by mimicking learned stylistic features or by exploiting gaps in the classifier’s representation.
Future research could explore the adversarial robustness of authorship validation classifiers by applying evasion techniques (e.g., prompt-engineered LLM outputs) and studying model responses.
Such work could also investigate defenses such as ensemble models, adversarial training, or detecting unnatural language patterns.

\textit{Other collaboration tools.}
Attacks similar to lateral spear phishing can be attempted through other collaboration tools (e.g., Microsoft Teams \cite{teams}, Salesforce Slack \cite{slack}, and Cisco Webex \cite{webex}).
Additional work is needed to adapt the techniques discussed in this paper to collaboration tools beyond email \cite{mimecast2023}.

\bibliographystyle{IEEEtran}
\bibliography{snn}

\appendix

\section {Email examples}

\subsection {Clean (authentic) email example} \label{sec: clean}
\begin{spverbatim}
Subject: A resume for Londom

This is a resume of one guy I met in Houston a few months ago.
He comes across as a very bright and  qualified individual. He is interested in a position in London. Who is the best person in London to forward the resume to?

Vince
\end{spverbatim}

\subsection {Generated (inauthentic) email example} \label{sec: generated}
\begin{spverbatim}
Subject: Recommending an Exceptional Candidate for Placement in London

Dear Colleague,

I am writing to recommend an impressive candidate that I had the good fortune to meet in Houston several months ago. This individual possesses a remarkable blend of intellect and qualifications that makes him stand out.

He expressed a keen interest in pursuing an opportunity in London. I am seeking your guidance in this regard - could you please advise who would be the ideal individual in London to whom I might forward his resume?

Best Regards,
Vince
\end{spverbatim}

\subsection {Char-CNN false positive example} \label{sec: fp}
\begin{spverbatim}
Subject: Re: Endorsement of a Stellar Baylor MBA Candidate for Summer Internship

Much appreciated,

Vince
\end{spverbatim}

\subsection {Char-CNN false negative example} \label{sec: fn}
\begin{spverbatim}
Subject: Risk management portal update

http://search.netscape.com/
cgi-bin/search?cp=nsiwidsrc&
search=reverse+search&x=36&y=1
\end{spverbatim}
\end{document}